\documentclass[aps,prd,preprint,nofootinbib]{revtex4}
\usepackage{graphicx,epstopdf}
\usepackage{amsmath,amssymb}
\usepackage{url}
\newcommand{\lsim}{\mathrel{\mathop{\kern 0pt \rlap
  {\raise.2ex\hbox{$<$}}}
  \lower.9ex\hbox{\kern-.190em $\sim$}}}
\newcommand{\gsim}{\mathrel{\mathop{\kern 0pt \rlap
  {\raise.2ex\hbox{$>$}}}
  \lower.9ex\hbox{\kern-.190em $\sim$}}}
\renewcommand{\thefootnote}{\fnsymbol{footnote}}
\renewcommand{\thefootnote}{\arabic{footnote}}

\begin{document}
\title{Hidden Sector Dirac Dark Matter, Stueckelberg $Z^\prime$ Model\\ 
and the CDMS and XENON Experiments}
\author{Kingman Cheung$^{1,2,3}$, Kuo-Hsing Tsao$^4$ and Tzu-Chiang Yuan$^5$}

\affiliation{
$^1$Division of Quantum Phases \& Devices, School of Physics,
Konkuk university, Seoul 143-701, Korea\\
$^2$ Physics Division, National Center for Theoretical Sciences, Hsinchu, 
 Taiwan\\
$^3$ Department of Physics, National Tsing Hua University,
Hsinchu, Taiwan \\
$^4$ Department of Physics, University of Illinois, Chicago, IL 60607, USA \\
$^5$ Institute of Physics, Academia Sinica, Nankang, Taipei 11925, Taiwan
}
\renewcommand{\thefootnote}{\arabic{footnote}}
\date{\today}        

\begin{abstract}

  For the milli-charged Dirac dark matter in the Stueckelberg $Z'$
  model, we discuss the contributions from the vector couplings of the
  dark matter with the neutral gauge bosons to the spin-independent
  scattering cross section in the direct detections.  We also compute
  the effective coupling between the fermionic dark matter particle
  and the standard model Higgs boson generated through a triangular
  loop of $Z$ and/or $Z'$ bosons which may contribute to the
  spin-independent scattering cross section at the quantum level.  We
  show that the latter contribution is consistent with the most recent
  experimental limits from CDMS II and XENON100.  
  In the case that the dark matter particle carries a
  milli-charge of order $O(10^{-3}e)$, we found that it would lose
  all its kinetic energy by colliding with nucleons in the atmosphere
  before reaching the detector. Even though we use
  the Stueckelberg $Z'$ model for illustration, the results we obtain
  are rather general and applicable to other $Z$-$Z'$ portal-type
  hidden-sector models as well.

\end{abstract}

\maketitle

\section{Introduction}

The presence of cold dark matter (CDM) in our Universe is now well
established by the very precise measurement of the cosmic microwave
background radiation in the Wilkinson Microwave Anisotropy Probe
(WMAP) experiment \cite{wmap7}.  One of the most appealing and natural
CDM particle candidates is provided by supersymmetric models with
$R$-parity conservation \cite{hooper}.  This $R$-parity conservation
ensures the stability of the lightest supersymmetric particle (LSP) so
that the LSP can be CDM.  The LSP is in general the lightest
neutralino, a linear combination of neutral electroweak (EW) gauginos
and Higgsinos.  Since the nature of LSP depends on its precise
compositions, its detection can vary a lot.

One of the proposed methods of detecting the dark matter is through 
direct search experiments. 
The dark matter particles move at a velocity relative to the detecting
materials.  It will recoil against the nucleons, and create either (1) 
phonon, (2) scintillation or (3) ionization signal, 
which can be amplified by electronics.  The CDMS experiment
is of the first type, while the XENON experiment is combination of types
(2) and (3). Recently, the CDMS II finalized their search in Ref. \cite{cdms}.
When they opened the black box in their blind analysis,
they found two candidate events, which are
consistent with background fluctuation at a probability level of
about 23\%.  
Nevertheless, the signal is not conclusive.  The CDMS then improves 
upon the upper limit on the 
spin-independent cross section $\sigma^{\rm SI}_{\chi N}$ to
$3.8 \times 10^{-44} \; {\rm cm}^2$ for $m_\chi \approx 70 $ GeV.  
Using this new limit we have put a new bound on the
Higgs-dark-matter coupling \cite{ours},
which can be implied to an upper limit on
the Higgs boson invisible width\footnote
{There have been a few works \cite{ours,others} on the Higgs invisible decay
in light of the CDMS II result.}.
More recently, XENON100 \cite{xenon100} has also reported no signals in their detectors
and put a similar upper limit on the 
spin-independent cross section $\sigma^{\rm SI}_{\chi N}$ to
$3.4 \times 10^{-44} \; {\rm cm}^2$ for $m_\chi \approx 55 $ GeV. 

There is a new class of models for dark matter candidates, motivated
by hidden-sector models. It could be a fermion or boson inside
the hidden sector.  The interactions with the standard model (SM) particles are only
possible via $Z$-$Z'$ mixing, Stueckelberg-type mixing, Higgs-portal
type models.  We focus on the former two possibilities in the
following.  Two of us \cite{st-dark} and Nath {\it et al.} \cite{nath} had
proposed the Dirac fermion, denoted by $\chi$, in the hidden sector as
the dark matter candidate.  Correct relic density can be obtained by
adjusting the mass of the dark matter, the mass of the $Z'$ boson, and
the coupling strength \cite{st-dark}.  
In this work, we study the effective coupling between 
the dark matter and the nucleon in the Stueckelberg $Z'$ model.
There are two possibilities for the effective interactions:
(I) The general mixings among the photon, $Z$ and $Z'$ make it possible
for the hidden sector dark matter couples to the SM quarks at the tree level;
(II) It is obvious that the fermion
$\chi$ has no direct coupling to the Higgs boson at the tree level.
However, we point out that an effective
coupling $g_{h\chi\chi}$ can be generated between the Dirac fermion
$\chi$ and the Higgs boson through a triangular loop of $Z$ and/or
$Z'$ bosons.  It therefore gives rise to spin-independent scattering
cross section that may be measurable with direct detection
experiments.  
We show that the effective couplings in case (I) will provide useful constraints
in the model, while case (II) will give rise to spin-independent cross sections, 
which are consistent with the most recent CDMS II and XENON100 results.

The organization of the paper is as follows.  
In the next section, we briefly describe
the spin-independent cross sections in terms of the effective couplings
between the dark matter and the nucleon for both vector and scalar contributions
in a model independent manner.
In Sec. III, we summarize some major features of the Stueckelberg $Z'$ model
and the generic $Z-Z'$ mixing model.
In Sec. IV, we calculate the effective couplings for both vector and scalar contributions 
in the Stueckelberg $Z'$ model.
Explicit results for the scalar contributions are presented for the Stueckelberg $Z'$ model as well as 
the generic $Z$-$Z'$ mixing model. Comparisons with the experimental limits are made.
Finally, we summarize in Sec. V.

\section{Direct Detection}

Direct detection of the fermionic dark matter depends on the assumption how dark matter interacts 
with the nuclei or with the quarks from the more fundamental viewpoint.
Contributions to the spin-independent cross section depend on the underlying mechanism 
is either dominated by vector gauge boson exchange or Higgs boson exchange 
or combination of both, depending on the specific model.
We will first discuss these contributions in a model independent way.

Suppose the effective interactions between the dark matter particle 
(denoted by $\chi$ in the following) and the quarks are given by
\begin{equation}
{\cal L }_{\rm eff} = \sum_{q} \large\{ \alpha_q^S \,\overline{\chi} \chi \, \bar q q +
            \alpha_q^V \,\overline{\chi} \gamma^\mu \chi \, \bar q \gamma_\mu q \large\} \;\; ,
\end{equation}
with the model dependent scalar and vector couplings specified by 
$\alpha_q^S$ and $\alpha_q^V$ respectively,
then the spin-independent cross section between $\chi$ 
and each of the nucleon (taking the average between proton and neutron)
is given by, (assuming a Dirac type fermion)
\begin{equation}
\label{111}
\sigma^{\rm SI}_{\chi N} = \frac{\mu_{\chi N}^2}{\pi}\; \left (
 \left|G^N_s \right |^2 + \frac{|b_N|^2}{256} \right )\;,
\end{equation}
where $\mu_{\chi N} = m_\chi m_N / ( m_\chi + m_N)$ is the reduced mass
between the dark matter particle and the nucleon $N = ( p, n) $, and 
\begin{eqnarray}
G^N_s & = & \sum_{q} \langle N | \bar q q | N \rangle \;
  \alpha^S_q 
 \; = \; \sum_{q} \alpha_q^S\, f^N_{Tq} \frac{ m_N}{m_q} \;, \\
 & = & \sum_{q=u,d,s} \alpha_q^S\, f^N_{Tq} \frac{ m_N}{m_q} 
 + \frac{2}{27} f^N_{Tg} \sum_{q=c,b,t} \alpha_q^S\, \frac{ m_N}{m_q}  \; ,
\end{eqnarray}
where the relation $\langle N | \bar q q | N \rangle = f^N_{Tq} m_N/m_q$ has been used for the 
the nucleon matrix element $\langle N | \bar q q | N \rangle$.

While the expression for the vector contribution $b_N$ of a {\em whole nucleus} $(A,Z)$ is
$b_N \equiv \alpha_u^V (A+Z) + \alpha^V_d (2A-Z)$, we take the 
average between proton and neutron (assume number of protons is 
about the same as neutrons in the nuclei) and thus obtain the expression
for a single nucleon 
\begin{equation}
b_N = \frac{3}{2} \left( \alpha_u^V + \alpha_d^V \right ) \;\; .
\end{equation}
This is useful for direct comparison with the results given by experiments,
where usually the dark-matter-nucleon cross sections are reported.

In the case of the dominance by Higgs boson exchange, 
we write $\alpha_q^S$ as 
$$\alpha_q^S = - \frac{  g_{h\chi \chi} g_{h q q } }{ m^2_h }$$ 
and 
$G_s^N$ takes the form
\begin{equation}
G^N_s = \sum_{q=u,d,s} f^N_{Tq} \frac{ m_N}{m_q} \,
\left ( - \frac{  g_{h\chi \chi} g_{h q q } }{ m^2_h } \right  ) 
+
\frac{2}{27} f^N_{Tg} \sum_{q=c,b,t}  \frac{ m_N}{m_q} \,
\left ( - \frac{  g_{h\chi \chi} g_{h q q } }{ m^2_h } \right  ) \;\; ,
\end{equation}
where $g_{h\chi\chi}$ is the effective coupling between the dark matter
particle $\chi$ and the Higgs boson, $g_{hqq}$ is the Yukawa coupling
of the quark $q$.
Note that the $m_q$ dependence in the Yukawa coupling $g_{h qq}$ 
will be cancelled by the $1/m_q$ dependence coming from the matrix element
$\langle N | \bar q q | N \rangle $.
In the scenario where the Higgs boson exchange is dominant, 
the experimental upper limit on the spin-independent cross section can imply an upper limit
for the dark matter-Higgs coupling, which is more or less model
independent.

Default values of the parameters used, e.g. in DarkSUSY \cite{darksusy}
are
\begin{eqnarray}
&&f^p_{Tu} = 0.023,\;\; f^p_{Td} = 0.034,\;\; f^p_{Ts} = 0.14,\;\;
f^p_{Tg} = 1 - f^p_{Tu} - f^p_{Td } -  f^p_{Ts} =  0.803 \;, \nonumber \\
&&f^n_{Tu} = 0.019,\;\; f^n_{Td} = 0.041,\;\; f^n_{Ts} = 0.14,\;\;
f^n_{Tg} = 1 - f^n_{Tu} - f^n_{Td } -  f^n_{Ts} =  0.8 \; .
\end{eqnarray}
We take the average between proton and neutron for $\sigma^{\rm SI}_{\chi N}$.
%%
%Thus the average value of $G^N_s$ is
%\begin{equation}
%- G^N_s \simeq g_{h\chi\chi} \frac{g m_N}{2 m_W} \frac{1}{m^2_h} (0.3766) \;.
%\end{equation}
For $m_\chi \sim O(100)$ GeV, $\mu_{\chi N} \approx m_N$ and the
spin-independent cross section was estimated to be \cite{ours}
\begin{equation}
\sigma^{\rm SI}_{\chi N} \approx \frac{ g^2 m_N^4}{4 \pi m_W^2}\, \frac{1}{m_h^4}\,
 g^2_{h\chi \chi} (0.3766)^2 \;.
\end{equation}
Using the new CDMS II limit of $\sigma^{\rm SI}_{\chi N} < 3.8 \times 10^{-44} \;
{\rm cm}^2$ and taking $m_h = 120$ GeV,  
we can obtain an upper limit on the Higgs-dark-matter coupling \cite{ours}
\begin{equation}
\label{limit}
g^2_{h\chi \chi} \alt 0.04 \; \; .
\end{equation}
Similar constraint on $g^2_{h\chi \chi}$ can be deduced by using the XENON100 limit.

\section{Models}

\subsection{Stueckelberg $Z'$ model}

The details of the Stueckelberg $Z'$ model can be found in
Ref. \cite{nath-st}, and more specifically the couplings used in this
study can be found in Ref. \cite{ww-z}.  Here we give a brief account.
The Stueckelberg extension of the SM (StSM) \cite{nath-st} is
obtained by adding \cite{st-dark,nath} a hidden sector associated with an extra $U(1)_C$
interaction, under which the SM particles are neutral.  Assuming there
is no kinetic mixing between the two $U(1)$'s, the Lagrangian
describing the system is given by
$$
{\cal L} ={\cal L}_{\rm SM} + {\cal L}_{\rm StSM}
$$
with
\begin{eqnarray}
{\cal L}_{\rm SM} &=& -\; \frac{1}{4} W_{\mu\nu}^a \, W^{a\mu\nu}
          - \frac{1}{4} B_{\mu\nu}\, B^{\mu\nu} + i \bar f \gamma^\mu D_\mu f
            + D_\mu \Phi^\dagger \, D^\mu \Phi - V(\Phi^\dagger \,\Phi) \; , \\
{\cal L}_{\rm StSM} &=&
         - \frac{1}{4} C_{\mu\nu}\, C^{\mu\nu}
 + \frac{1}{2} \left( \partial_\mu \sigma + M_1 C_\mu + M_2 B_\mu \right )^2
  +  {\overline \chi} \left( i \gamma^\mu D^X_\mu - M_\chi \right) \chi ~,
\label{stusm}
\end{eqnarray}
where $W_{\mu\nu}^a(a=1,2,3)$, $B_{\mu\nu}$ and $C_{\mu\nu}$ are the
field strength tensors of the gauge fields $W_\mu^a$, $B_\mu$, and
$C_\mu$, respectively; $f$ denotes a SM fermion, while $\chi$ is a
Dirac fermion in the hidden sector which may play a role as
milli-charged dark matter in the Universe
\cite{st-dark,nath} and $M_\chi$ is its mass; $\Phi$
is the SM Higgs doublet; and $\sigma$ is the Stueckelberg axion
scalar.  The covariant derivatives $D_\mu = (\partial_\mu + i g_2 \vec
T \cdot {\vec W}_\mu + i g_Y\frac{Y}{2} B_\mu )$ and $D^X_\mu =
(\partial_\mu + i g_X Q^\chi_X C_\mu )$.

After the electroweak symmetry breaking of 
$\langle \Phi \rangle = v/\sqrt{2}$ with a vacuum expectation
value $v \simeq 246$ GeV, the mass term for 
$V \equiv ( C_\mu,\, B_\mu,\, W^3_\mu )^T$ is given by
\begin{equation}
- \frac{1}{2} V^{\mathrm T}  M_{\mathrm{Stu}}^2  V  \equiv   -\frac{1}{2}
  \left ( C_\mu,\, B_\mu,\, W^3_\mu \right ) \;
 \left( \begin{array}{ccc}
  M_1^2 & M_1 M_2 & 0 \\
 M_1 M_2 & M_2^2  + \frac{1}{4} g_Y^2 v^2 & -\frac{1}{4} g_2 g_Y v^2 \\
 0 & -\frac{1}{4} g_2 g_Y v^2  & \frac{1}{4} g_2^2 v^2 \end{array} \right ) \;
 \left( \begin{array}{c}
       C_\mu \\
       B_\mu \\
       W^3_\mu \end{array} \right ) \;.
\end{equation}
One can easily show that the determinant of $M_{\mathrm{Stu}}^2$ is
zero, indicating the existence of at least one zero eigenvalue to be
identified as the photon mass.  A similarity transformation $O$ can bring
the mass matrix $M^2_{\mathrm{Stu}}$ into a diagonal form
\begin{equation}
\left( \begin{array}{c}
                                        C_\mu \\
                                        B_\mu \\
                                        W^3_\mu \end{array} \right ) =
  O \, \left( \begin{array}{c}
                                        Z'_\mu \\
                                        Z_\mu \\
                                        A_\mu \end{array} \right )
\;\; , \;\;\;\;\;
 O^{\mathrm T}  M_{\mathrm{Stu}}^2 \, O \, =
 \mathrm{Diag} \left[ M^2_{Z'},\, M^2_Z,\, 0 \right]  \; .
 \label{rotate}
\end{equation}
Explicit formulas for the matrix elements of $O$ in terms of the fundamental parameters
in the Lagrangian can be found in Refs. \cite{st-dark, nath, nath-st, ww-z}. 
The couplings between the neutral gauge bosons and the Higgs are given by
\begin{eqnarray}
\label{HVVCouplings}
{\cal L}_{{\rm Higgs}-Z-Z^\prime} & = &
\frac{1}{8} \left(H^2 + 2 v H \right)
\left[
\left( g_2 O_{32} - g_Y O_{22}  \right)^2 Z_\mu Z^\mu +
 \left( g_2 O_{31} - g_Y O_{21} \right)^2 Z^\prime_\mu Z^{\prime\mu} \right. 
\nonumber \\
&+ & \left. 2 \left( g_2 O_{31} - g_Y O_{21}  \right) 
\left( g_2 O_{32} - g_Y O_{22}  \right)  Z_\mu Z^{\prime\mu} \right] \; .
\end{eqnarray}
The neutral current interactions are given by
\begin{eqnarray}
- {\cal L}^{NC}_{\rm int} &=& \bar f\, \gamma^\mu \left[
  \left( \epsilon_{Z'}^{f_{L}} P_L + \epsilon_{Z'}^{f_{R}} P_R\right)\, Z'_\mu
+ \left( \epsilon_{Z}^{f_{L}} P_L + \epsilon_{Z}^{f_{R}}P_R \right)\, Z_\mu
+ e Q_{\rm em} A_\mu \right ] f \nonumber\\
&+&
 {\overline \chi} \gamma^\mu \left[
 \epsilon^\chi_{Z'} Z'_\mu 
 +\epsilon^\chi_Z Z_\mu
 +  \epsilon^\chi_\gamma A_\mu
 \right ]\, \chi \; ,
\label{chiral1}
\end{eqnarray}
with
\begin{eqnarray}
 \epsilon_Z^{f_L} &=& \frac{g_2}{\cos\theta} \cos\psi
 \left[ \left( 1 -  \epsilon \sin \theta \tan \psi \right) T^3_f
 - \sin^2\theta \left( 1 -  \epsilon \csc \theta \tan \psi \right) Q_f \right] \; , \nonumber\\
 \epsilon_Z^{f_R} &=& - \frac{g_2}{\cos\theta} \cos\psi
 \sin^2\theta \left( 1 -  \epsilon \csc \theta \tan \psi \right) Q_f \; , \nonumber\\
 \epsilon_{Z'}^{f_L} &=& - \frac{g_2}{\cos\theta} \cos\psi
 \left[ \left( \tan \psi +  \epsilon \sin \theta \right) T^3_f
 - \sin^2\theta \left(  \epsilon \csc \theta + \tan \psi \right) Q_f \right] \; , \nonumber\\
 \epsilon_{Z'}^{f_R} &=& \frac{g_2}{\cos\theta} \cos\psi
  \sin^2\theta \left(  \epsilon \csc \theta + \tan \psi \right) Q_f \;, \nonumber \\
  \epsilon^\chi_{V_i} & = & g_X Q^\chi_{X} O_{1i} \; ,
 \label{chiral2}
\end{eqnarray}
and $\epsilon \equiv \tan \phi$.
Therefore, the couplings among the Higgs boson, $Z$ and $Z'$ are
\begin{eqnarray}
C_{Z'Z'} & = & \left( g_2 O_{31} - g_Y O_{21} \right)^2 O_{11}^2 \;\; , \\
C_{ZZ} & = &  \left( g_2 O_{32} - g_Y O_{22} \right)^2 O_{12}^2 \;\; , \\
C_{Z'Z} & = & C_{ZZ'} = \sqrt{ C_{Z'Z'} C_{ZZ} } \;\; ,\nonumber \\
& = & \left( g_2 O_{31} - g_Y O_{21} \right) \left( g_2 O_{32} - g_Y O_{22} 
\right) O_{11} O_{12} \;\; .
\end{eqnarray}

With all these couplings and inputs we are ready to compute the
effective coupling $g_{h\chi\chi}$ and thus the spin-independent
cross section $\sigma^{\rm SI}_{\chi N}$.

\subsection{$Z$-$Z'$ mixing models}

Before mixing the neutral current interactions are
\begin{equation}
- {\cal L}_{\rm NC} = \frac{g_2}{\cos\theta_{\rm w} } \, \sum_f\, \bar f \,
\gamma^\mu\, ( g_v - g_a \gamma_5 ) \, f \; Z_{1\mu} 
 + g_X Q_X^\chi \overline \chi \gamma^\mu \chi \; Z_{2\mu}
\end{equation}
where $Z_1$ and $Z_2$ are the unmixed states, which are then rotated into
mass eigenstates $Z,Z'$ via a mixing
\begin{equation}
 \left(  \begin{array}{c} 
            Z_1 \\
            Z_2  \end{array} \right ) = \left( \begin{array}{cc}
                 \cos\theta  & - \sin\theta \\
                 \sin \theta  & \cos\theta  \end{array} \right )\;
        \left( \begin{array}{c} 
                     Z \\
                     Z' \end{array} \right ) \;\; .
\end{equation}
The  neutral current gauge interactions of the hidden fermion $\chi$ in this model are then
\begin{equation}
 -   {\cal L}^{\chi}_{\rm NC} = g_X Q_X^\chi 
 \left( \cos\theta\, Z'_\mu + \sin\theta \, Z_\mu \right ) \overline{\chi} \gamma^\mu \chi \;\; .
\end{equation}
The interactions among the Higgs boson and the $Z,Z'$ are
\begin{equation}
- {\cal L}_{hZZ} = \frac{1}{8} \frac{g_2^2}{\cos^2\theta_{\rm w} }
 (H^2 + 2 v H)  \;  \left( \cos\theta\, Z_\mu - \sin\theta \, Z'_\mu \right )^2 \;\; 
\end{equation}
where $\theta_{\rm w}$ is the Weinberg's angle.
Therefore, the coefficients $C$'s in this model are given by
\begin{eqnarray}
C_{Z'Z'} & = & \frac{g_2^2}{\cos^2\theta_{\rm w}}  \; \sin^2\theta \; 
        \cos^2 \theta \;\;, \\
C_{ZZ} & = &  \frac{g_2^2}{\cos^2\theta_{\rm w}}  \; \cos^2\theta \; 
    \sin^2 \theta \;\;, \\
C_{Z'Z} & =& C_{Z Z'}= -
   \frac{g_2^2}{\cos^2\theta_{\rm w}} \; \sin^2\theta \cos^2\theta  \;\; .
\end{eqnarray}

\section{Effective Couplings in Stueckelberg $Z'$ model}

\subsection{Tree-level mixing}

We note that the dark matter $\chi$ in the present Stueckelberg $Z'$
model is assumed to be a Dirac particle. Thus it has vector couplings
to all the neutral gauge bosons $\gamma$, $Z$ and $Z'$. In particular,
it carries milli-charged while couples to photon \cite{st-dark,
  nath}. There are also axial couplings among the dark matter with $Z$
and $Z'$, but they only contribute to the spin-dependent cross
sections.  For multi-component dark matter model of Stueckelberg type
with both Dirac and Majorana hidden fermions, see the recent work of
Ref.\cite{multi-comp-dm}.

Due to the $t$-channel pole, one might expect the photon exchange diagram will be the dominant contribution 
for the spin-independent dark matter-nucleon cross section and thus the 
CDMS II result can be used to eliminate the milli-charged dark matter model. 
However, for the milli-charged dark matter to get detected inside the underground detectors, 
it has to traverse our whole atmosphere and penetrate through the surface rock. 
The milli-charged dark matter interacting with the ordinary matter 
through the long range 
Coulombic force might lose all its kinetic 
energy before it reaches the detector.
Following the analysis in \cite{foot}, 
one can estimate the stopping distance to be
\begin{eqnarray}
L & \approx & \frac{m_A^2 m_\chi v_\chi^4}
{8 \pi \rho \left(\tilde\epsilon^\chi_\gamma \alpha Z\right)^2 \log\left( E_R^{\rm max} / E_R^{\rm min} 
\right)} \nonumber \\
& \approx & 0.27 
\left( \frac{10^{-3}}{\tilde \epsilon^\chi_\gamma} \right)^2
\left( \frac{m_A}{32 \, {\rm GeV}} \right)^2
\left( \frac{16}{Z} \right)^2
\left( \frac{m_\chi}{100 \, {\rm GeV}} \right)
\left( \frac{5 \, {\rm g/cm^3}}{\rho} \right)
\left( \frac{v_\chi}{300 \, {\rm km/s}} \right)^4 \;\; [{\rm m}]
\end{eqnarray}
Here, $m_A$ is the mass of the ordinary matter with atomic number $Z$ and density $\rho$,
$\tilde\epsilon^\chi_\gamma = \epsilon^\chi_\gamma / e$ is the milli-charge of
the dark matter in unit of $e$, and $E_R^{\rm min, max}$ are 
the minimum and maximum recoil energies of the matter, respectively.
Following \cite{foot}, we set $\log\left( E_R^{\rm max} / E_R^{\rm min} 
\right) \sim 10$ to obtain the above estimation. 
This small stopping distance suggests milli-charged dark matter will not be able to reach the 
underground detectors unless $\tilde\epsilon^\chi_\gamma$ is very small of the order of
$10^{-8}$. Such kind of dark matter can arise in models with kinetic mixing. In the present
context of Stueckelberg $Z'$ model, $\tilde\epsilon^\chi_\gamma$ can be 
considerably larger, typically of the size of $10^{-3}$ \cite{st-dark,nath}.

For completeness, we also present the contributions from 
the $Z$ and $Z'$ diagrams.  The spin-independent dark matter-nucleon 
cross section from their vector couplings is given by the second term of Eq.~(\ref{111}),
\begin{equation}
\label{spin-indep-xsec-Z}
\sigma^{\rm SI}_{\chi N} = \frac{\mu^2_{\chi N} |b_N|^2}{256 \pi} \;\; ,
\end{equation}
where $b_N = \frac{3}{2} (\alpha_u^V + \alpha_d^V)$ assuming the number of
protons and the number of neutrons are about the same and taking
the average between proton and neutron.
For the $Z$ and $Z'$ couplings in the Stueckelberg $Z'$ model described above, we have
\begin{eqnarray}
\alpha_f^V & = & \frac{\epsilon^\chi_Z}{2 m_Z^2} 
\left( \epsilon^{f_L}_Z + \epsilon^{f_R}_Z \right) +
\frac{\epsilon^\chi_{Z'}}{2 m_{Z'}^2} 
\left( \epsilon^{f_L}_{Z'} + \epsilon^{f_R}_{Z'} \right) \; ,
\end{eqnarray}
for $f = u$ or $d$-quark.  For sufficiently heavy $Z'$, its contribution 
can be ignored.  With $\tan \phi = 0.01$ and $m_{Z'} = 1 $ TeV, 
we obtain an estimate for  
\begin{equation}
\sigma^{\rm SI}_{\chi N} \approx \,  0.44 \times 10^{-45} \;\; [{\rm cm}^2] \; .
\end{equation}
This is roughly two orders
 of magnitude below the current CDMS II and XENON100 limits.
 Thus, one can not put stringent constraint on the vector couplings of
 the dark matter $\chi$ with the $Z$ and $Z'$ bosons in the
 Stueckelberg model yet.  However, the long range Coulomb interaction
 between the milli-charged dark matter and the nuclei of the detectors
 may be too severe for the dark matter to reach the detector in this
 scenario.  In this case the above formula (\ref{spin-indep-xsec-Z})
 for the spin-independent cross section derived from the effective
 4-fermion local operator is no longer applicable.  On the other hand,
 the CDMS II and XENON100 limits can be used to place useful
 constraint on the mixing angle in generic $Z$-$Z'$ mixing model
 presented in previous section.  Furthermore, XENON100 is expected to
 improve its upper limit by one to two order of magnitudes in the
 future. The parameter space of the Stueckelberg $Z'$ model will be
 probed more effectively by this future improvement.

\subsection{Effective Higgs-Dark Matter Coupling}

The effective coupling between the SM Higgs and the hidden
milli-charged dark matter $\chi$ can be induced at one-loop
level. The calculation is similar to the one loop electroweak 
correction to the $H b \bar b$ coupling except there are 
no $W^\pm$ and unphysical Higgs bosons $G^{\pm,0}$ 
running inside the loops.
Thus to simplify our calculation we will proceed using the 
't Hooft - Feynman gauge.

We can write down the following amplitude for the effective coupling of 
Higgs-$\chi$-$\chi$
\begin{equation}
\overline \chi(p') \left[ F(q^2)  + i \gamma_5 G(q^2) \right]\chi(p) \; H
\end{equation}
with $F(q^2)$ and $G(q^2)$ being the scalar and pseudoscalar form factors, 
and $q = p - p'$.  For on-shell $\chi$ in the initial and final states, 
we find that $G(q^2) = 0$ and 
\begin{equation}
F(q^2) =  - \frac{ \left( g_X Q^\chi_X \right)^2 }{8 \pi^2} \frac{m_W m_\chi}{g_2}
\sum_{(i,j) = \left\{ (Z',Z'),(Z,Z),(Z',Z),(Z,Z') \right\} }  C_{ij}
 \int_0^1 dx \int_0^x dy  
\frac{\left( 1 + y \right)}{\Delta_{ij} (x,y) }
\end{equation}
where $g_X Q^\chi_X$ is the gauge coupling of the fermion $\chi$ to the
vector gauge boson in the hidden sector, 
\begin{equation}
\Delta_{ij} (x,y)  = 
y^2  m^2_\chi  + ( x - y) m^2_i + ( 1 - x) m^2_j  - ( 1 - x) ( x - y ) q^2 - i 0^+
\end{equation}
and $C_{Z_i Z_j}$ contains the coupling of $H Z_i Z_j$ and the mixing 
angles of $Z_i$ and $Z_j$ with the hidden-sector dark matter.
We will give these couplings explicitly in
the Stueckelberg $Z'$ model and the generic $Z$-$Z'$ mixing 
model in the subsections.
The effective Higgs-dark-matter coupling $g_{h\chi\chi}$ relevant
for spin-independent cross section is then given by
\begin{equation}
g_{h\chi\chi} = F(q^2 = 0)
\end{equation}
in which the elastic scattering of the $\chi$ is at $q^2 \approx 0$.

If kinematically allowed, the SM Higgs can decay into the invisible $\overline \chi \chi$ 
mode and its width is given by 
\begin{equation}
\label{inv-width}
\Gamma \left( h \longrightarrow \overline \chi \chi \right) = 
\frac{m_h}{8 \pi} \vert F \left( m_h^2 \right) \vert^2 \left( 1 - 
\frac{4 m_\chi^2}{m_h^2} \right)^{\frac{3}{2}}  \; \; .
\end{equation}
%

%%%%%%%%%%%%%%%%%%%%%%%%%%%%%%%%%%%%%
\begin{figure}[t!]
\centering
\includegraphics[width=5.5in]{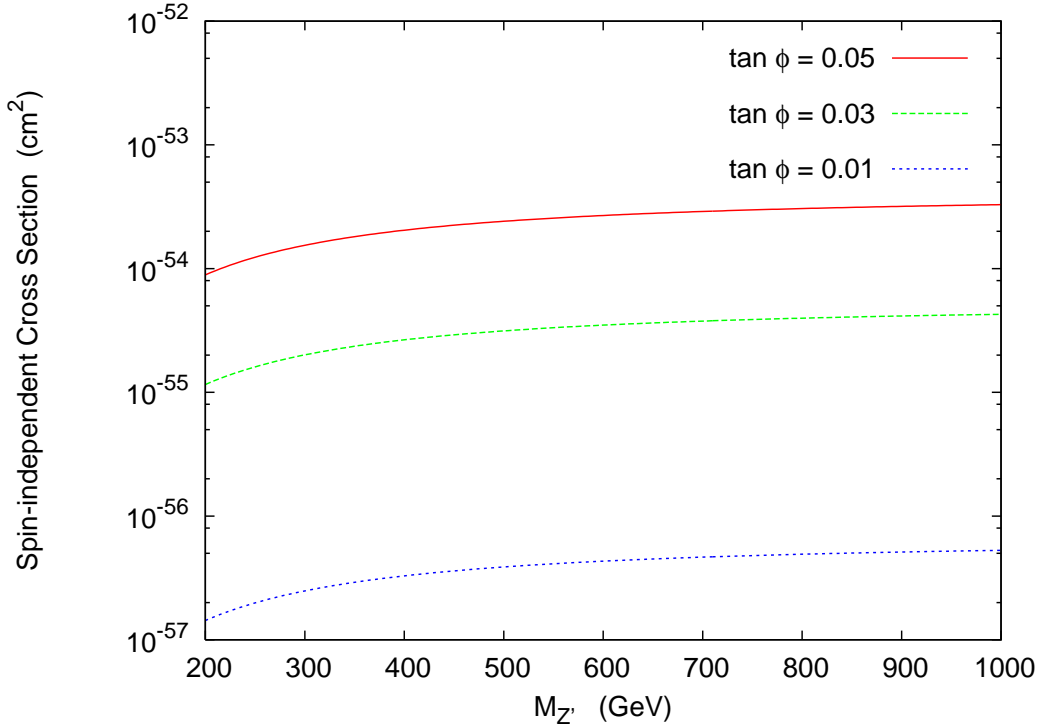}
\caption{\small \label{fig1}
The spin-independent cross section $\sigma^{\rm SI}_{\chi N}$ versus
the mass of the $Z'$ boson in the Stueckelberg model.  Inputs are
$g_X Q_X^\chi = g_2$, $m_\chi = 100$ GeV and $m_h = 120$ GeV.}
\end{figure}
%%%%%%%%%%%%%%%%%%%%%%%%%%%%%%%%%%%%%

With $\tan\phi = M_2 / M_1$ and $m_{Z'}$ fixed, 
all parameters of the Stueckelberg $Z'$ model are fixed. 
In Fig. \ref{fig1}, we show the $\sigma^{\rm SI}_{\chi N}$ versus
the mass of $Z'$ with the inputs: $g_X Q_X^\chi = g_2$, 
$m_\chi = 100$ GeV, $m_h = 120$ GeV and three different values of 
$\tan\phi$ = 0.01, 0.03 and 0.05. 
It is evident from this plot that for a wide range of parameters\footnote{
The spin-independent cross section can get a significant boost 
in the large $\tan \phi$ scenario discussed in \cite{ww-z}. However, the 
results are still several orders below the current CDMS II limit.}
in this model,
the predictions of the spin-independent cross sections are consistent 
with the latest CDMS II experiment.
No severe constraints can be deduced for this model from the current 
CDMS experiment.
Similar conclusions can be obtained for the generic $Z$-$Z'$ mixing model since
the mixing angle is generally constrained to be rather tiny $(\alt 10^{-3})$.

We have also checked that the invisible decay width of the Higgs decay 
$H \to {\overline \chi} \chi$ given by Eq.(\ref{inv-width}) 
is very tiny and will not have any significant 
effects on the Higgs boson decay.

\section{Summary}

In summary, we have calculated the nucleon-dark-matter scattering
due to (i) tree-level mixing between the hidden sector gauge boson and the
SM gauge bosons, and (ii)  the one-loop induced coupling between the
SM Higgs boson and the Dirac fermionic dark matter in the hidden sector.
Such couplings can contribute
to the spin-independent cross section for the dark matter direct
detection. 
However, for a wide class of $Z$-$Z'$ models we found that
these scalar contributions to
the spin-independent cross sections are well below 
the current CDMS II and XENON100 experiments.
The current experimental limits do not provide useful constraints for 
the vector couplings of the 
dark matter with the neutral gauge bosons in the Stueckelberg $Z'$ 
model but may place 
more restrictive constraints in generic $Z$-$Z'$ mixing models.
Furthermore, in the case that the dark matter particle
has a milli-charge of order $O(10^{-3}e)$ (e.g. in the Stueckelberg-type
mixing), it cannot traverse the atmosphere to the detector because
it lost all its kinetic energy by colliding with nucleons. In such a 
case, the direct detection limits cannot apply.
Projected upper limit on the spin-independent cross section for XENON100 
is expected to have improvements by one to two orders of magnitudes 
and thus it can
probe the parameter space of the model more effectively.

%\newpage

\section*{Acknowledgments}
We would like to thank M. Drees, D. Feldman, P. Nath and O. Seto for useful communications.
The work was supported in parts by the National Science Council of
Taiwan under Grant Nos. 96-2628-M-007-002-MY3 and
98-2112-M-001-014-MY3, the National Center for Theoretical Sciences (NCTS), 
the Boost Project of NTHU, and 
the WCU program through the KOSEF funded by the MEST (R31-2008-000-10057-0).

\end{document}